\RequirePackage{fix-cm}
\documentclass[twocolumn,epjc3]{svjour3}  
\smartqed  
\RequirePackage{graphicx}
\journalname{Eur. Phys. J. C}
\usepackage{cite}
\usepackage{amsmath}
\usepackage{amsfonts}
\usepackage{amssymb}
\usepackage{makeidx}
\usepackage{graphicx}
\usepackage{lmodern}
\usepackage{color}

\begin{document}
\title{\bf 
	Perturbative approach to {\boldmath $f(R)$}-gravitation in  
	FLRW cosmology} 
%
\author{Pham Van  Ky\thanksref{e1}
        \and
Nguyen Thi Hong  Van\thanksref{e2}
        \and 
Nguyen Anh Ky\thanksref{e3}
}
\thankstext{e1}{e-mails: phamkyvatly AT gmail.com (co-first author)}
\thankstext{e2}{e-mail: nhvan AT iop.vast.vn (co-first author)} 
\thankstext{e3}{e-mail: anhky AT iop.vast.vn (corresponding author)}
\institute{\small 
\textit{Mathematical-, high energy- and astro-physics group, CTP,}\\
\small
\textit{Institute of physics},\\ 
\small
\textit{Vietnam academy of science and technology (VAST) },\\ 
\small
\textit{10 Dao Tan, Ba Dinh, Hanoi, Viet Nam.}
}
\date{Received: date / Accepted: date}
%
%
\maketitle
\begin{abstract}
\noindent

The $f(R)$ theory of gravitation developed perturbatively around the general theory of relativity with cosmological constant (the \text{$\Lambda$}CDM model) in a flat FLWR geometry is considered. As a result, a general explicit cosmological  solution that can be used for any model with an arbitrary, but well-defined, $f(R)$ function (just satisfying given  perturbation conditions) is derived. This  perturbative solution shows how the Hubble parameter $H (t)$ depends on  time (along with the cosmological constant and the matter density) to adapt to  the evolution of the Universe. To illustrate, this approach is applied to some specific test models.
One of these models appears to be more realistic as it could describe three phases of the Universe's evolution.

Despite the fact that the perturbation is applied for a flat FLWR geometry  (according to the current cosmological observation) indicates that the obtained solution can mainly describe the evolution of the late Universe, it may also work for an early Universe. As a next step, the present method can be applied to the case with a more general FLRW geometry to increase the precision of the description of different stages in the evolution of the Universe.

 Finally, it is shown that in a desription of the Universe's evolution the  perturbative $f(R)$-theory can be considered as an effective GR with the cosmological constant 
$\Lambda$ replaced by an effective parameter $ \Lambda_{eff}[\rho(t)] $. This trick leads to a simpler way of solving an $f(R)$-theory regardless its specific form.
\end{abstract}
\section{Introduction}
\noindent 

Published by A. Einstein in 1915, the general theory of relativity (GR) \cite{Wein, Land} has become one of the pillars of modern physics and is very well tested both theoretically and experimentally. This great 
 theory can explain or predict different astrophysical and cosmological phenomena, such as those in the radiation-dominated and the matter-dominated eras of the Universe (see, for example, \cite{Penrose:1964wq,Gillessen:2008qv,LIGOScientific:2017vwq,LIGOScientific:2016aoc,Kramer:2021jcw}, for some recent results supporting the GR) 
but it fails to explain other phenomena, mainly in the acceleration periods of the Universe. Moreover, so  successful but the GR is still facing problems of quantum gravity \cite{rovelli,Linde:1990flp}, the cosmic inflation \cite{Starobinsky:1979ty,Linde:1990flp,Liddle:2000cg, Guth:1980zm,Guth:1997wk, Mukhanov:2005sc}, the origin of dark matter \cite{Arbey:2021gdg} (see also, for example, \cite{Drewes:2016upu} in a particle physics aspect), the accelerated expansion of the Universe (called otherwise the dark energy problem) \cite{SupernovaSearchTeam:1998fmf,SupernovaCosmologyProject:1998vns,Crease:2009zz}, etc. These problems call for an extension or modification of the GR but so far there has been no satisfactory theory suggested.
To name a few famous theories extending the GR. The string theory (for a review, see, for instance, \cite{Polchinski:1998rq}) expected as a "\textit{theory of everything}" is a very complicated  ultra-high energy theory and weakly developed in the phenomenological aspect, therefore, it is very hard for an experimental test. The similar situation is  with a more special theory - supergravity \cite{VanNieuwenhuizen:1981ae,Tanii:2014gaa}. Moreover, these theories are based on the concept of supersymmetry \cite{West:1990tg} of which the LHC and other experiments have not found any sign so far.
In general, to explain the formation and the evolution of the Universe is always a challenge in physics calling for extended theories. One of the first attempts to modify the GR was made by A. Einstein himself not long  after the birth of this theory.\\

Einstein modified his original GR by adding a term called the cosmological constant \cite{{Weinberg:1988cp},Peebles:2002gy} to the Lagrangian of the Einstein-Hilbert action. By introducing the cosmological constant, Einstein tried to "keep" the Universe static as it could have been generally imagined in his epoch. Later, Einstein, however, blamed himself that the introduction of the cosmological constant was a blunder after knowing the theoretical works by A. Friedmann \cite{Friedman:1922kd} and G. Lema\^itre \cite{Lemaitre:1931zza} and the observation by E. Hubble \cite{Hubble:1929ig} proving the expansion of the Universe. According to these works the Universe is not static at all but dynamical, more precisely, as stated by Hubble's law, different parts  of the Universe, e.g. galaxies, are receding from each other at a speed proportional to their relative distance. Moreover, today we know that the Universe is not only expanding\ \cite{Hubble:1929ig} but is expanding at an increasing rate  \cite{SupernovaSearchTeam:1998fmf,SupernovaCosmologyProject:1998vns,Crease:2009zz}. Despite this fact, the idea of the cosmological constant is not completely useless. It has been expected to help us in solving the dark energy problem by treating the cosmological constant proportional (or equivalent) to the dark energy- or vacuum energy density (corresponding to an equation of state with negative pressure) as we can see by examining Einstein's equation with cosmological constant  directly. Nevertheless, another difficulty arising here is this treatment gives rise to the so-called ``\textit{cosmological constant problem}" or "\textit{vacuum catastrophe}": there is a huge discrepancy, of many orders of magnitude, between the observed value of the vacuum energy density (generated by the cosmological constant) and that estimated by quantum field theory \cite{Weinberg:1988cp,Peebles:2002gy,Martin:2012bt,Adler:1995vd}. It is the ever known biggest discrepancy between theory and experiment/observation in physics. That means that the QFT approach suffers from a serious ailment, not to mention its complexity.  
The next problem is, with the cosmological constant in the Einstein equation the Hubble parameter $H$ is also a constant (independent of time) and therefore, the GR does not accurately describe the evolution of $H$ over time (as very slowly but  $H$ changes over time). Adding the cosmological constant is just the first modification of the GR. 
Combined with the cold dark matter (CDM) issue, it leads to the $\Lambda$CDM model, which, however, is lately shown to be not very accurate as the observed Universe is less "clumpy" than predicted by this model \cite{DES:2022urg}. Therefore, it is not the final word on a cosmological model. Since the deviation is not much, the realistic model should not differ too much from the $\Lambda$CDM one.
There have been so far many other attempts to extend or modify the GR in order to explain phenomena beyond 
the GR. Among them we choose the so-called $f(R)$ theory of gravitation \cite{DeFelice:2010aj,thomas,Capozziello:2011et,Nojiri:2010wj}, or just the $f(R)$ theory or  $f(R)$-gravitation, for short, which is one of the simplest modified theories of the GR, but expected by us to solve the problems emerging from the very geometry of the space-time.\\

Nowadays, the $ f(R) $ theory, due to its relative simplicity, to a great extent, has become a hot topical issue and has attracted  much attention of a number of astrophysicists and  cosmologists (see, for instance,  \cite{Linde:1990flp, Mukhanov:2005sc, DeFelice:2010aj, Nojiri:2010wj, thomas,  Capozziello:2011et,Odintsov:2019evb,Nojiri:2019fft,Nojiri:2006gh} and references therein). Another advantage of  the $f(R)$ theory is that, the latter through a (conformal)  transformation to an alternative form - the scalar-tensor theory \cite{DeFelice:2010aj,thomas} (in particular, the Jordan-Brans-Dicke theory \cite{Wein}), can make communication between particle physics (QFT) and cosmology. In this framework the scalar fields may play a crucial role and they, e.g., the inflatons\footnote{This idea of Alan Guth, however, was later rejected by himself \cite{Guth:1997wk} but still used elsewhere \cite{Jegerlehner:2014mua}.} could be sometimes \cite{Guth:1980zm} treated as the Higgs bosons \cite{Higgs:1964pj} believed now to be discovered ten years ago by the LHC collaborations ATLAS and CMS  \cite{ATLAS:2012yve,CMS:2012qbp} (see also \cite{AnhKy:2015fqw} for a review on the introduction, the search and the discovery of the Higgs boson). This conforms to the fact that different cosmological problems, such as those of cosmic inflation, dark matter, dark energy, etc., can be investigated in the viewpoints of both geometry and particle physics compatible with each other via, in particular, the $f(R)$ theory. Instead of the particle physics (-QFT) approach \cite{Linde:1990flp,DeFelice:2010aj}, here, within the $f(R)$ theory, we will follow the geometrical path to study some aspects of the evolution of the Universe.\\

Various  models based on the $f(R)$ theory have been suggested, but, so far, to our knowledge, most works have been devoted to models with $f(R)$ of specific forms rather than a general one. 
For example, two of popular versions (see, for instance, also \cite{Nojiri:2003ft}) of this theory are the model with $f(R)=R+\alpha R^2$ (with $\alpha$ being a coefficient independent of $R$ and other curvature quantities) used to explain the accelerated expansion of the Universe during the inflationary era \cite{DeFelice:2010aj,Starobinsky:1980te}, and the model with $f(R )=R+\beta/R$ (with the coefficient $\beta$ independent of $R$ and other curvature quantities) used to explain the accelerated expansion of the Universe in the late epochs including the present one (the dark energy problem) \cite{DeFelice:2010aj}, etc. A combination of the models of these two types can make a good model for both the early and the late Universe (for some other models, see, for example, \cite{Ky:2019gbj} and references therein). Each $f(R)$ model can explain some phenomena in some period of the Universe's evolution, but none of them is  perfect and powerful enough to describe different phenomena beyond the GR in different cosmic times.  To our knowledge, it is still difficult to conclude which is the right model and in most cases (see, for example, recent work   \cite{Nojiri:2019fft}) they are combined with the introduction of additional, usually scalar (or pseudo-scalar), fields such as axions and axion-like ones when the problem of the dark matter is also incorporated in a non-geometric way which we don't follow here. This makes the matters more cumbersome because of problems with quantum field theory and it is not clear yet if the dark matter has a particle physics origin. Instead, because of its visual  gravitational effect we prefer to treat the dark matter as the geometric background of space-time or other geometric origin (see, e.g.,  \cite{P.Ky,Boehmer:2007kx}). Some models considered in \cite{Nojiri:2006gh} are claimed to be realistic but they seem to be compatible only with the Solar System and cosmology without a cosmological constant. Regardless any model is considered, to solve the corresponding (extended) Einstein equation occupies a central position.\\

In general, to find a solution, especially, an exact one, of an $f(R)$-theory is very hard, even, sometimes, impossible. To simplify the situation, we can use a perturbation method and impose some reasonable conditions such as the  spherical symmetry which is a good approximation in many cases including that of a homogeneous and isotropic Universe according to the cosmological principles. Perturbative solutions of the $f(R)$ theory in a central field of a distinct gravitational source  (such as a star, a black hole, etc.) and their implications were studied recently in Refs. \cite{AnhKy:2018pbb, Ky:2019gbj, VanKy:2020xxj}. An application of this approach to gravitational radiation in the $f(R)$-theory is also underway \cite{fR-gw}.  This  approach, however,  has not yet been applied to the Universe as a whole. In this article, we go ahead to look for an approximate solution of a general $f(R)$ theory (with an arbitrary, but well-defined, $f(R)$ function) applied to a homogeneous and isotropic Universe, 
adopting, thus, the Friedmann-Lema\^itre-Robertson-Walker (FLRW) metric \cite{Weinberg:2008zzc}. We make perturbation around the GR theory (always with the  cosmological constant included, unless otherwise stated),  
which means that the $f(R)$ theories used are slightly different from the GR 
(as the GR has been very precisely tested, it is assumed that any deviation from this theory should be small). As a result, we obtain an explicit perturbative solution for an FLRW Universe. This perturbative solution improves the GR in the sense that
the Hubble parameter H is not a constant at all but depends on time to accommodate the evolution of the Universe. As the observed  Universe is almost flat \cite{Adler:2005mn}, 
it is reasonable to work here, in a perturbation approach, with a flat FLRW metric (for the late Universe, at least). 
 At the meantime, we are working on the early Universe 
where the flat metric might not be a good approximation but the same approach could be applied to a general (non-flat) FLRW metric to describe a curved Universe. Here it is worth mentioning that recent Planck's observations showed that the Universe could be closed, i.e., curved with a positive, albeit very small, curvature  \cite{DiValentino:2019qzk} but this statement requires confirmation by  further observations.\\

In this article the following conventions are used:
\begin{itemize}	
	\item Signature of the Minkowski metric: $ (+, -, -, -) $, that
	is, the infinitesimal distance is given as ($x^0=ct$) 
	$$ds^2=\eta_{\mu\nu}dx^{\mu}dx^{\nu} \equiv {dx^0}^2-dx^2-dy^2-dz^2.$$
	
	\item Riemannian curvature tensor:
	\begin{align*}
	R^{\alpha}_{~\mu\beta\nu}=&\frac{\partial \Gamma^\alpha_{\mu\beta}}{\partial x^\nu} - \frac{\partial \Gamma^\alpha_{\mu\nu}}{\partial x^\beta} + \Gamma^\alpha_{\sigma\nu}\Gamma^\sigma_{\mu\beta} -  \Gamma^\alpha_{\sigma\beta}\Gamma^\sigma_{\mu\nu}.
	\end{align*}
	
	\item Ricci tensor: $ R_{\mu\nu}=R^\alpha_{~\mu\alpha\nu}$.
	
	\item Scalar curvature: $ R=g^{\mu\nu}R_{\mu\nu},~ g^{\mu\nu}=g^{\mu\nu}(x) $.
	
	\item Energy-momentum tensor of a macroscopic object: 
	\begin{equation*}
	T_{\mu\nu}= \frac{1}{c^2}(\varepsilon + P)u_\mu u_\nu - Pg_{\mu\nu},
	\end{equation*}
	where $ u^\mu = \displaystyle \frac{dx^\mu}{d\tau}=c\frac{dx^\mu}{ds} $, while $ \varepsilon  $ and $ P $ are the energy density and the pressure, respectively. 
\end{itemize}
The plan of the present paper is the following. The next section is devoted to the search for  perturbative solutions of a general $f(R)$-theory, before applying them to specific models in Sect. 3. Some numerical discussions are made in Sect. 4. The effective cosmological constant issue is discussed in Sect.5, while Sect. 6 is designed for concluding remarks and briefly outlines further research. 
%
\section{Perturbative solutions of $f(R)$--theory of gravitation and FLRW cosmology}
\noindent

The Einstein equation of the GR (or the GR equation for short) with the cosmological constant $\Lambda $ included can be derived from the Lagrangian 
$${\cal L}_\Lambda=R-2\Lambda . $$
This (the GR with the cosmological constant, or more generally, the  \text{$\Lambda$}CDM model) is the theory around which our perturbative $f(R)$-theory will be developed. The $f(R)$-theory is     
a more-general theory with the Lagrangian  ${\cal L}_G= f(R)$, where $f(R)$ is a scalar function of the scalar curvature $R$, leading to the following equation generalizing the Einstein equation  \cite{DeFelice:2010aj, thomas, Capozziello:2011et}:
%
	\begin{align}
	& f'(R)R_{\mu\nu}-g_{\mu\nu}\square f'(R)+\nabla_{\mu}\nabla_{\nu} f'(R)-\frac{1}{2}f(R)g_{\mu\nu}
\nonumber\\ & 
	=-kT_{\mu\nu}, \label{v1}
	\end{align}
%
where $k= \displaystyle \frac{8\pi G}{c^4} $ , $\square = \nabla_{\mu}\nabla^{\mu} $ with  $\nabla_{\mu}$ the covariant derivative, and $ f'(R)=\displaystyle\frac{d}{dR}f(R) $.  This theory in an appropriate condition, as shown below, can describe the evolution of the Universe in different stages. To this end we will work with those $f(R)\equiv {\cal L}_G$ which can be developed perturbatively around ${\cal L}_\Lambda$. Since our Universe in large scale is homogeneous, isotropic and nearly flat, it makes sense to choose its geometry based on a flat (or almost flat) FLRW metric.\\

Using the flat  FLRW metric (here the unit in which $c=1$ is used) \cite{Wein, DeFelice:2010aj}
\begin{align}
ds^2=dt^2-a^2(t)\left( dx^2 + dy^2 + dz^2 \right), \label{a39}
\end{align}
we find the following non-zero Ricci tensor elements \cite{Wein}
\begin{align}
R_{00}&=3\frac{\ddot{a}}{a} \label{a40},\\ 
R_{ij}&=\left( \frac{\ddot{a}}{a}+2\frac{\dot{a}^2}{a^2}\right)g_{ij}, \label{a41}
\end{align}
where $ i,j=1,2,3 $, 
therefore, 
\begin{align}
R=6\left( \frac{\ddot{a}}{a}+\frac{\dot{a}^2}{a^2}\right). \label{a44}
\end{align}
With denoting
\begin{align}
&T^0_{~0}=\rho, \label{vtf1}\\
&T^1_{~1}=T^2_{~2}=T^3_{~3}=-P. \label{vtf2}
\end{align}
and using \eqref{a39}, the  equation \eqref{v1} leads to two independent equations of the Universe, 
\begin{equation}
3f'(R)\frac{\ddot{a}}{a}-3\frac{\dot{a}}{a}\dot{f}'(R)-\frac{1}{2}f(R)=-k\rho, \label{vtf3}
\end{equation}
\begin{equation}
f'(R)\left(\frac{\ddot{a}}{a}+2\frac{\dot{a}^2}{a^2} \right) -2\frac{\dot{a}}{a}\dot{f}'(R)-\ddot{f}'(R)-\frac{1}{2}f(R)=kP, \label{vtf4}
\end{equation}
where $ \displaystyle \dot{f}'(R)=\frac{\partial}{\partial t}f'(R) $ and $ \displaystyle  \ddot{f}'(R)=\frac{\partial^2}{\partial t^2}f'(R) $. 
Next, for later use in our perturbation approach we write  $f(R)$ in the form 
\begin{equation}
f(R)=R-2\Lambda+\lambda h(R), \label{vtf5}
\end{equation}
with $ \Lambda $ and $ \lambda $  being constants,  and $ h(R) $ being a scalar function of $ R $  such that $| \lambda h(R)|\ll |R-2\Lambda|$. 
%
\setlength{\parindent}{0pt}
Thus, taking  \eqref{a44} and \eqref{vtf5} into account we rewrite 
\eqref{vtf3} and \eqref{vtf4}
in the following way
%
	\begin{align}
	3\lambda h'(R)\frac{\ddot{a}}{a}-3\frac{\dot{a}^2}{a^2}-3\lambda\frac{\dot{a}}{a}\dot{h}'(R)
	+\Lambda-\frac{\lambda}{2}h(R)=-k\rho, \label{k3}
	\end{align}
%
and 
	\begin{align}
	& [-2+\lambda h'(R)]\frac{\ddot{a}}{a}+[-1+2\lambda h'(R)]\frac{\dot{a}^2}{a^2}  -2\lambda\frac{\dot{a}}{a}\dot{h}'(R)
\nonumber\\ & 
	-\lambda\ddot{h}'(R)
	+\Lambda-\frac{\lambda}{2}h(R)=kP, \label{k4}
	\end{align}
%
respectively, and then, combining the latter equations  \eqref{k3} and \eqref{k4} we obtain 
%
	\begin{align}
	&6[1-\lambda^2 h'^2(R)]\frac{\dot{a}^2}{a^2}  +3\lambda \dot{h}'(R)[2+\lambda h'(R)]\frac{\dot{a}}{a}
    \nonumber\\& 
    -2\Lambda [1+\lambda h'(R)]+3\lambda^2\ddot{h}'(R)h'(R)+3\lambda kh'(R)P
    \nonumber\\&
     +\lambda h(R)[1+\lambda h'(R)]-k\rho [2-\lambda h'(R)]	
    =0. \label{k5}
	\end{align}
%
For the GR (when $\lambda=0$) these equations are reduced to    
\begin{align}
R^\mu_{~\nu}-\frac{1}{2}R\delta^\mu_{~\nu}+\Lambda \delta^\mu_{~\nu}=-kT^\mu_{~\nu}, \label{k7}
\end{align}
\begin{equation}
R_{GR}=4\Lambda + kT=4\Lambda +k(\rho-3P)=\overline{\Lambda}, \label{k8}
\end{equation}
\begin{equation}
H_{GR}=\frac{\dot{a}}{a}=\sqrt{\frac{\Lambda }{3}+\frac{k\rho}{3}}, \label{k14h}
\end{equation}
where the notations $T=T^\mu_\mu$ and 
$\overline{\Lambda}=4\Lambda +k(\rho-3P)$ are used. \\

\setlength{\parindent}{6pt}

Now, suppose that in the formula \eqref{vtf5} we consider the term $ \lambda h(R) $ as a perturbation term  about the GR (when $f(R)={\cal L}_\Lambda \equiv R-2\Lambda $). Thus in the equations  \eqref{k5} we can treat the terms  containing $ \lambda $ as perturbation terms  about the GR equations. We solve the equation \eqref{k5} by a perturbation method as follows:  substituting the solutions of the GR equations \eqref{k8}  and \eqref{k14h}  into the perturbative terms of \eqref{k5}, we  get  
%
	\begin{align}
	& 6\frac{\dot{a}^2}{a^2} -6\lambda^2 h'^2(\overline{\Lambda})\left( \frac{\Lambda }{3}+\frac{k\rho}{3}\right)  
   \nonumber\\ &
    +3\lambda \dot{h}'(\overline{\Lambda})[2+\lambda h'(\overline{\Lambda})]\sqrt{\frac{\Lambda }{3}+\frac{k\rho}{3}}  -2\Lambda [1+\lambda h'(\overline{\Lambda})] 
\nonumber\\ &   
	 +3\lambda^2\ddot{h}'(\overline{\Lambda})h'(\overline{\Lambda}) +\lambda h(\overline{\Lambda})[1+\lambda h'(\overline{\Lambda})]
\nonumber\\& 
	-k\rho [2-\lambda h'(\overline{\Lambda})]
	+3\lambda kh'(\overline{\Lambda})P=0. \label{k9b}
	\end{align}
%
That means we solve Eq. \eqref{k5} at its first order of perturbation \eqref{k9b}. 
From \eqref{k9b} we can immediately obtain the Hubble parameter $ H(t)\equiv \dfrac{\dot{a}}{a} $, with 
%
	\begin{align}
H^2(t)&=\dfrac{\Lambda}{3} [1+\lambda h'(\overline{\Lambda})]+\frac{1}{6}k\rho [2-\lambda h'(\overline{\Lambda})]
\nonumber\\
 &
 -\frac{1}{2}\lambda \dot{h}'(\overline{\Lambda})[2+\lambda h'(\overline{\Lambda})]\sqrt{\frac{\Lambda }{3}+\frac{k\rho}{3}}
 \nonumber\\&
 +\lambda^2 h'^2(\overline{\Lambda})\left( \frac{\Lambda }{3}+\frac{k\rho}{3}\right)
   -\frac{1}{2}\lambda^2\ddot{h}'(\overline{\Lambda})h'(\overline{\Lambda})
\nonumber\\ 
&	-\dfrac{1}{6}\lambda h(\overline{\Lambda})[1+\lambda h'(\overline{\Lambda})]-\dfrac{1}{2}\lambda kh'(\overline{\Lambda})P. \label{H2}
	\end{align}
	Note that from the equation of state $P=\omega \rho $ for matter ($\omega=0$) we have $ P=0$, thus,  formula \eqref{H2} becomes
	\begin{align}
	H^2(t)&=\dfrac{\Lambda}{3} [1+\lambda h'(\overline{\Lambda})]+\frac{1}{6}k\rho [2-\lambda h'(\overline{\Lambda})]
\nonumber\\&  
    -\frac{1}{2}\lambda \dot{h}'(\overline{\Lambda})[2+\lambda h'(\overline{\Lambda})]\sqrt{\frac{\Lambda }{3}+\frac{k\rho}{3}}
\nonumber\\&
    +\lambda^2 h'^2(\overline{\Lambda})\left( \frac{\Lambda }{3}+\frac{k\rho}{3}\right)
    -\frac{1}{2}\lambda^2\ddot{h}'(\overline{\Lambda})h'(\overline{\Lambda}) 
    \nonumber\\& 
	-\dfrac{1}{6}\lambda h(\overline{\Lambda})[1+\lambda h'(\overline{\Lambda})], \label{H3}
	\end{align}
where $\overline{\Lambda}$ now takes the value  
\begin{equation}
\overline{\Lambda}=4\Lambda +k\rho, \label{H4}
\end{equation}
and 
\begin{align}
&\dot{h}'(\overline{\Lambda})=\dot{\overline{\Lambda}}h''(\overline{\Lambda}), \label{dh1}\\
&\ddot{h}'(\overline{\Lambda})={\dot{\overline{\Lambda}}}^2h'''(\overline{\Lambda})+\ddot{\overline{\Lambda}}h''(\overline{\Lambda}). \label{dh2}
\end{align}
It is easily to find $a(t)$ via the formula 
\begin{equation}
a(t)=a_0 \exp\left[\int^t_{t_0}H(t)dt \right], \label{a1}
\end{equation}
with the constant $ a_0 =a(t_0) $ and $ H(t) $ given in \eqref{H3}.
Along with \eqref{k9b}, combining \eqref{k3} and \eqref{k4} we also have the equation  
\begin{align}
& [6+3\lambda h'(R)]\frac{\ddot{a}}{a}-[3+6\lambda h'(R)]\frac{\dot{a}^2}{a^2} +3\lambda\ddot{h}'(R)-\Lambda
\nonumber\\& 
+ \frac{\lambda}{2}h(R)+2k\rho +3kP=0, \label{gt}
\end{align}
which will be used later on. 
The solution \eqref{a1} with $ H(t) $ given in \eqref{H3} is perturbative and can be applied to any model of  the $f(R)$-theory satisfying perturbation conditions (see \cite{AnhKy:2018pbb} and below). Now let us do more detailed calculations for some specific models of the $f(R)$-theory.
%
\section{Applications to specific models}
\noindent

Let us apply the general results obtained above to some specific models (see also, for example, \cite{Nojiri:2003ft,Nojiri:2006gh} and references therein). 
\subsection{Model I: $ f(R)=R-2\Lambda + \lambda R^2 $}
\noindent

	This model resembles the Starobinsky model  \cite{Starobinsky:1980te} (but does not coincide with the latter corresponding to the Lagrangian  $f(R)_{Starobinsky}=R+R^2/6M^2$). Here $ h(R)=R^2 $, $ h'(R)=2R $ and $ h''(R)=2 $, thus, the formula \eqref{H3} now takes the form  
	%
		\begin{align}
		H^2(t)= ~ &\frac{\Lambda + k\rho(t)}{3}
	\nonumber\\& 
		- \frac{\lambda}{3}\left[ 4\Lambda+k\rho(t)\right]\left[16\lambda\Lambda^2 +8\lambda\Lambda k\rho(t) 
        \right.
		&\nonumber\\&\left. 
        +\frac{3}{2}k\rho(t)+\lambda k^2\rho^2(t)+6\lambda k\ddot{\rho}(t)\right] 
		\nonumber\\
		& +\frac{4\lambda^2}{3}\left[ \Lambda+k\rho(t)\right] \left[ 16\Lambda^2+8\Lambda k\rho(t)+k^2\rho^2(t)\right]
        \nonumber\\&
        - \lambda k\dot{\rho}(t)\left[ 2+8\lambda\Lambda+2\lambda k\rho(t)\right] \sqrt{\dfrac{\Lambda +k\rho(t)}{3}}. \label{hh1}
		\end{align}
	At $\lambda=0$, we obtain the GR value, namely,  
	\begin{equation}
	H^2_{GR}=\frac{\Lambda + k\rho(t)}{3}, \label{hh1a}
	\end{equation}
	which is the first term of \eqref{hh1}, while the rest terms are perturbative ones.  
	Since $ \rho(t)\longrightarrow 0 $ as  $ t\longrightarrow\infty $ 
	we get
	\begin{align}
	H(\infty)=H_{GR}(\infty)=\sqrt{\frac{\Lambda}{3}}. \label{hh2}
	\end{align}
	Up to the first order of $\lambda$,  the equation \eqref{hh1} has a relatively simple form  
		\begin{align}
		&H^2(t)=\frac{\Lambda + k\rho(t)}{3}-  \frac{\lambda k\rho(t)}{2}\left[ 4\Lambda+ k\rho(t)\right]
		\nonumber\\&
        - 2\lambda k\dot{\rho}(t)\sqrt{\dfrac{\Lambda +k\rho(t)}{3}}. \label{hh1b}
		\end{align}
	
	Let us now derive an approximate form of function $ \dot{\rho}(t)$ in \eqref{hh1b}. From the formula \eqref{v1}, we have the equation
	\begin{equation}
	\nabla_\mu T^\mu_{~\nu}=0. \label{k21a}
	\end{equation} 
	It follows that 
	\begin{equation}
	\dot{\rho}+3H(\rho+P)=0. \label{k21b}
	\end{equation}
	Taking $ H\simeq  \sqrt{\displaystyle\frac{\Lambda +k\rho(t)}{3}} $ 
	(and the equation of state $P=\omega \rho $ for matter $\omega=0$) into account we get  
	\begin{equation}
	\dot{\rho}(t)\simeq-\sqrt{3}\rho(t)\sqrt{\Lambda+k\rho(t)}, \label{ro1a}
	\end{equation}
	or
	\begin{equation}
	\ddot{\rho}(t)=\dfrac{3\rho(t)\left[ 2\Lambda+3k\rho(t)\right] }{2}. \label{ro1aa}
	\end{equation}
	Inserting \eqref{ro1a} in \eqref{hh1b} we get 
	(with the constant $c$ put back),
	\begin{equation}
	H(t)=\frac{\dot{a}}{a}=\sqrt{\frac{1}{3}c^2\Lambda +\frac{1}{3}c^4k\rho(t)+\frac{3\lambda}{2}c^6k^2\rho^2(t)}, \label{ro3}
	\end{equation}
	where $H(t) $ is the Hubble parameter at time $ t $. 
	This time dependence $H(t)$ is depicted via $H(\rho)$ in Fig.~1.
		\begin{figure}[h]
		\begin{center}
			\includegraphics[scale=0.43]{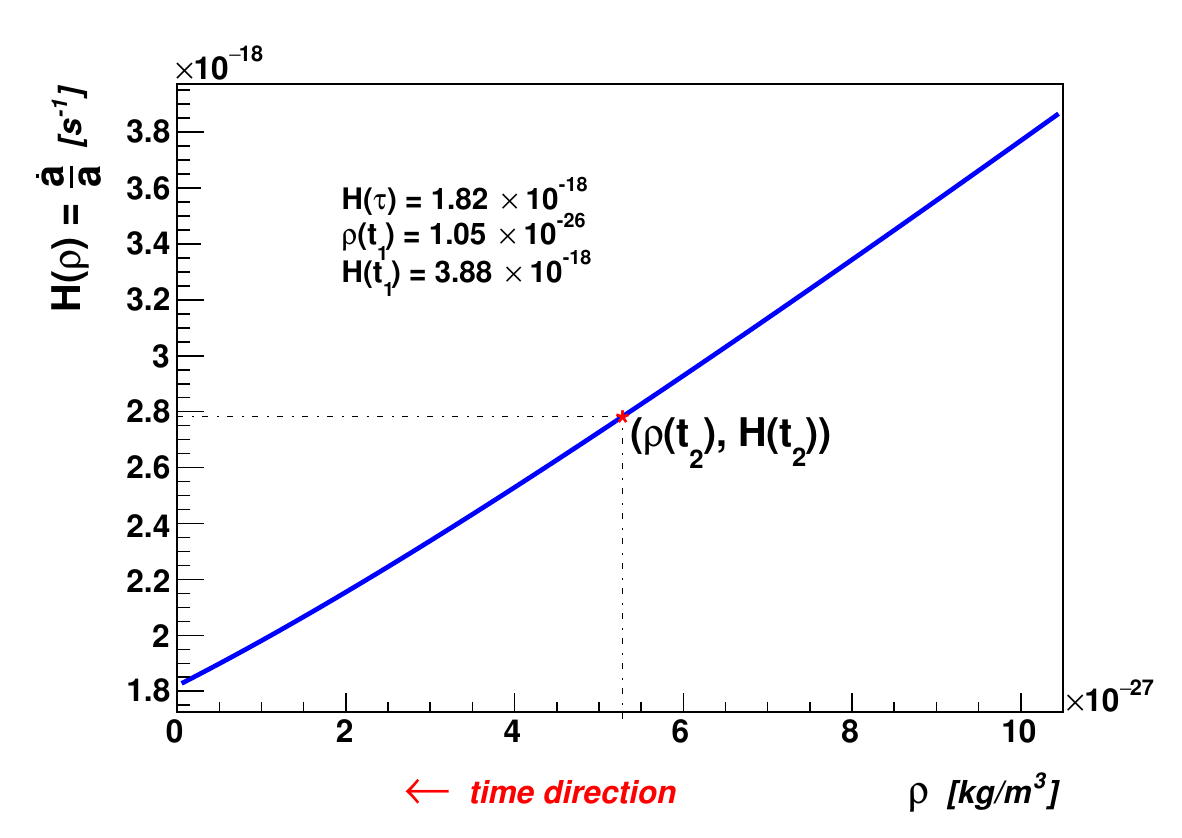}
			\caption{\label{fig:1}\textit{Hubble parameter $H$ as a function of $\rho$ between the time $t_1$, when the perturbation condition became applicable, and a very far future according to Eq. \eqref{ro3} in the Model-I.  The point $(\rho(t_2),H(t_2))$ is at the moment $t_2$ when the Universe's expansion starts to accelerate. The point $\rho=0$ is a limit in a distant future which may not be reachable. The time direction is from right to left.}}
		\end{center}
\label{Ht}		
	\end{figure}
Solving the equation \eqref{ro3} for $ a(t) $ we easily find
	\begin{equation}
	a(t)=a_0 \exp\left(\int^t_{t_0}\sqrt{\frac{1}{3}c^2\Lambda +\frac{1}{3}c^4k\rho(t)+\frac{3\lambda}{2}c^6k^2\rho^2(t)}dt \right), \label{ro4}
	\end{equation}
	where $ a_0=a(t_0) $. We note that according to \eqref{vtf5}  (with \eqref{k8} taken into account)  the perturbation condition is 
	\begin{equation}
	|\lambda h(R)| \ll 2\Lambda,  \label{dk1}
	\end{equation}
	which, for the model  $f(R)=R-2\Lambda + \lambda R^2 $ with $h(R)=R^2$, becomes
	\begin{equation}
	|\lambda| \ll \frac{1}{8\Lambda}. \label{dk2}
	\end{equation}
	As $ \Lambda\sim 10^{-52}m^{-2}$ {\color{red}\cite{Planck:2018vyg}} the perturbation upper bound of  
	$\lambda$ is very large, 
	$\lambda\ll 10^{51}m^2$. 
	With \eqref{ro1aa} and \eqref{ro3} the equation \eqref{gt} gets the form 
	\begin{align}
	&\left[ 1+4\lambda\Lambda+\lambda c^2k\rho(t)\right] \frac{\ddot{a}}{a}=
    \nonumber\\&
    \frac{c^2\Lambda}{3}-\frac{1+2\lambda\Lambda}{6}c^4k\rho(t)-\frac{19}{6}\lambda c^6k^2\rho^2(t)+\frac{4c^2\lambda\Lambda^2}{3}, \label{pttt1}
	\end{align}
This equation depicted in Fig.~2 shows how the Universe is expanding acceleratedly (as from the moment when the expansion acceleration becomes positive).
	\begin{figure}[h]
		\begin{center}
			\includegraphics[scale=0.43]{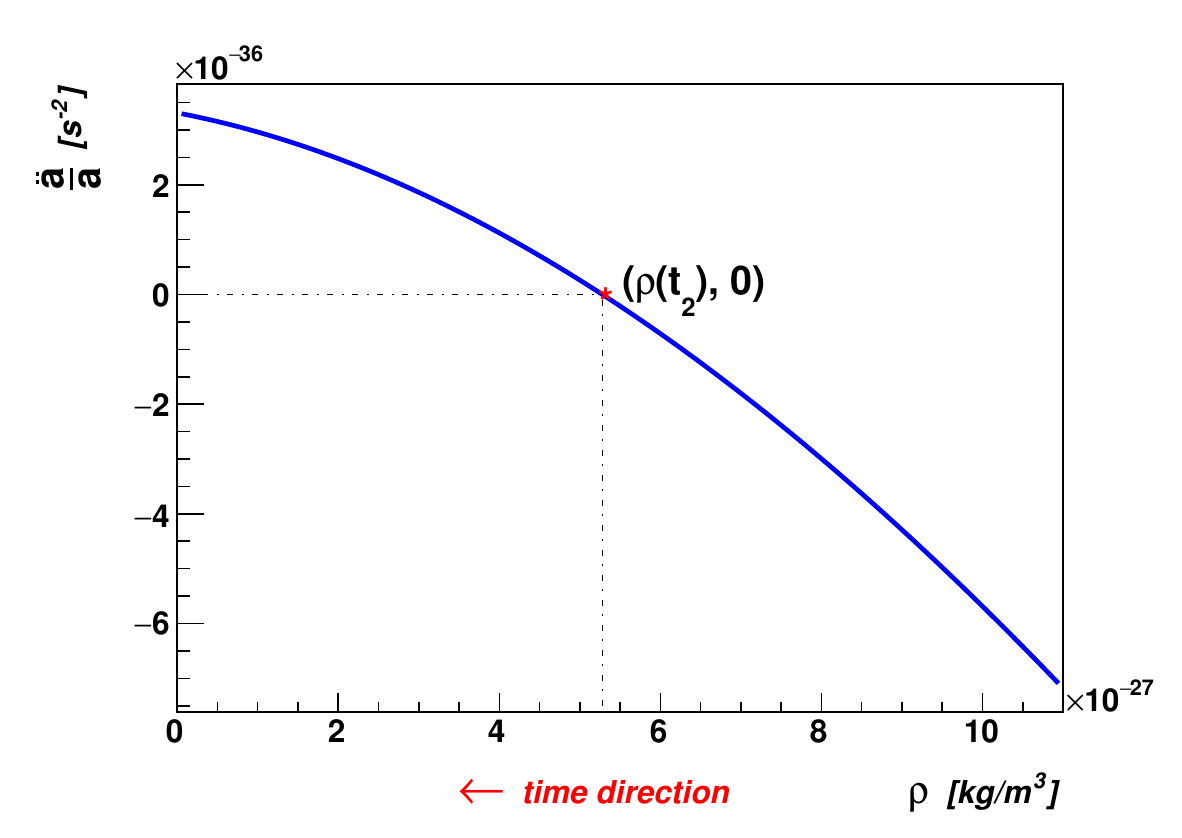}
\caption{\label{fig:2}\textit{Formula \eqref{pttt1} shows how the Univesre is acceleratedly exapanding in the Model-I. The point $(\rho(t_2),0)$ is at the moment $t_2$ when the acceleration of the Universe's expansion changes its sign (it is the transition point between two cosmic phases - the deceleration and the acceleration phases). The point $\rho=0$ is a limit which may never be reachable. The time direction is from right to left.}}
\end{center}
\label{Hdot}		
	\end{figure}
	From here we find the following condition for the accelerated expansion of the Universe (for $\ddot{a} > 0$):
\begin{equation}
c^2k\rho(t) < \frac{\sqrt{612\lambda^2\Lambda^2+156\lambda\Lambda+1}-1-2\lambda\Lambda}{38\lambda}. \label{pttt2}
\end{equation}
If $\lambda\Lambda\ll 1$, thus  $\sqrt{1+156\lambda\Lambda}\approx 1+78\lambda\Lambda$, the condition  \eqref{pttt2} becomes  
\begin{equation}
c^2k\rho(t) < 2\Lambda, \label{pttt2a}
\end{equation}
consistently with the GR.\\

	To finish this subsection, let us make some comments on \eqref{ro3}. In a late era, such as today's one, when the density $\rho$ is very small, the third term is dominated by the second term, while in an earlier era, such as the inflationary one, the third term dominates, therefore,   
	\begin{equation}
	H_{early}(t)=\frac{\dot{a}}{a}=\sqrt{\frac{1}{3}c^2\Lambda +\frac{3\lambda}{2}c^6k^2\rho^2(t)}, \label{ro5}
	\end{equation}
	thus, 
	\begin{equation}
	a_{early}(t)=a_0 \exp\left(\int^t_{t_0}\sqrt{\frac{1}{3}c^2\Lambda +\frac{3\lambda}{2}c^6k^2\rho^2(t)}dt \right). \label{ro6}
	\end{equation}
	Further discussions will be made in the next session.

\subsection{Model II: $f(R)=\eta R^{1+\varepsilon}-2\Lambda$}
\noindent 

In case $ \eta R^{1+\varepsilon}$ deffers from $R$ very little the function $f(R)$ of this model can be written in  the form 
$f(R)=R-2\Lambda + \lambda h(R)$  such that  
\begin{align}
\mid\lambda h(R)\mid \equiv \mid\eta R^{1+\varepsilon} - R\mid \ll \mid R-\Lambda\mid,  \label{mk2}
\end{align}
where the coefficient 
$\eta$ has the dimension $[\eta]=[\Lambda]^{-\varepsilon}$.  Depending on the sign of 
$1+\varepsilon$ the model could be more appropriate for an early or late Universe. 
This model modifies the model  with $f(R)=R^{1+\varepsilon}$ based on which the dark matter is investigated in  \cite{Boehmer:2007kx} (see also \cite{P.Ky}) in which 
\begin{align}
\varepsilon=\frac{v^2_{tg}}{c^2},\label{mk3}
\end{align}
where $v_{tg}$ is the tangential speed of the rotation of the edge of a galaxy. Here the cosmological constant is added to also resolve the dark energy problem.\\

\begin{figure}[h]
		\begin{center}
			\includegraphics[scale=0.43]{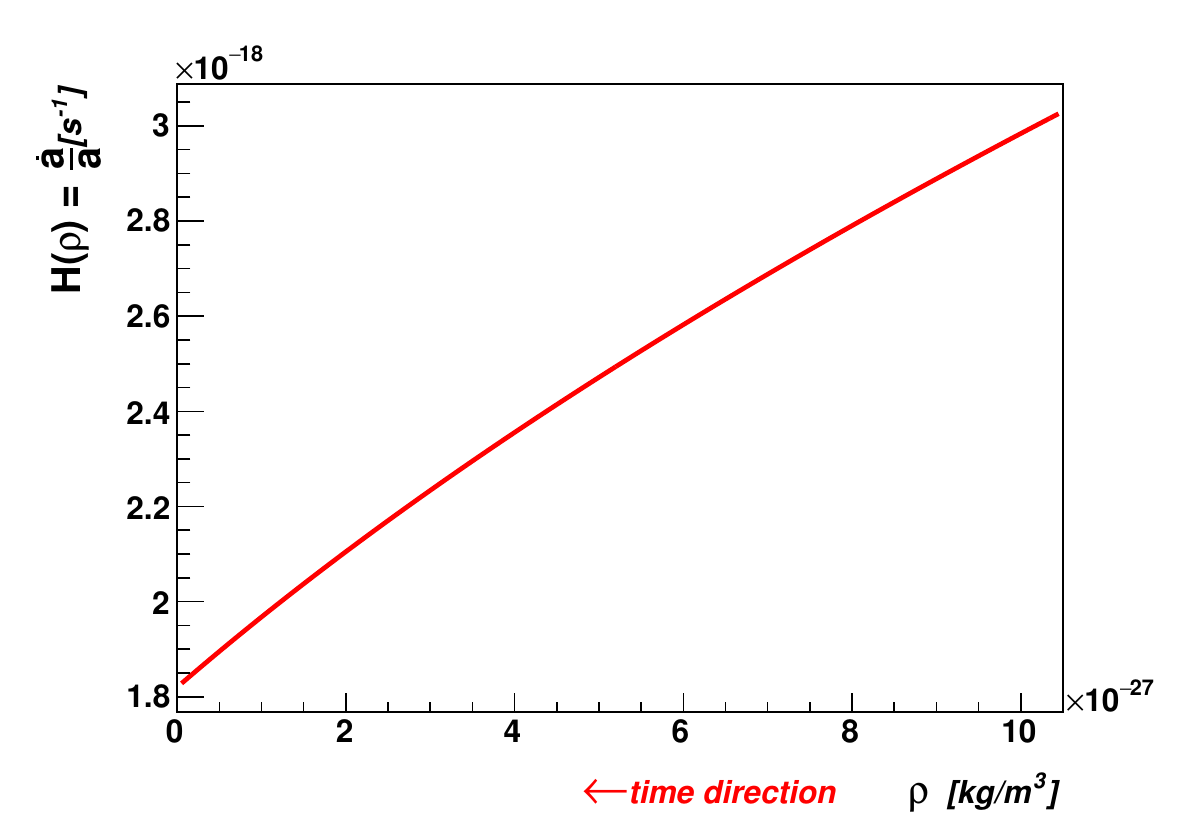}
\caption{\label{fig:3}\textit{The evolution of the Hubble parameter$H$
with respect to the matter density $\rho(t)$ in the model II. The time direction is from right to left.}}
\end{center}
\label{HdotIII}		
	\end{figure}

\begin{figure}[h]
		\begin{center}
			\includegraphics[scale=0.43]{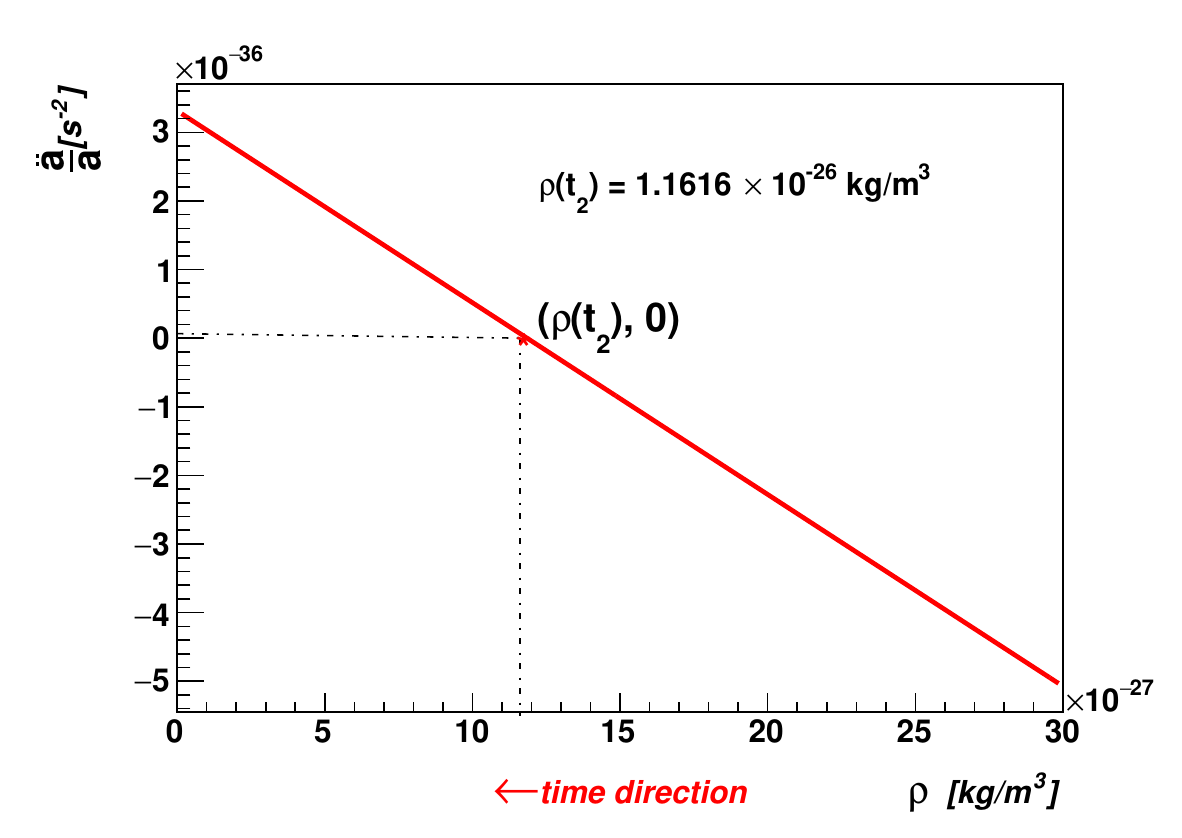}
\caption{\label{fig:4}\textit{Plot of  $\ddot{a}/a$ as a function of $\rho(t)$ in the model II. The point $(\rho(t_2),0)$ is at the moment $t_2$ when the acceleration of the Universe's expansion changes its sign. The point $\rho=0$ is a limit which may never be reachable. The time direction is from right to left.}}
\end{center}
\label{addotIII}		
	\end{figure}

\begin{figure}[h]
		\begin{center}
			\includegraphics[scale=0.43]{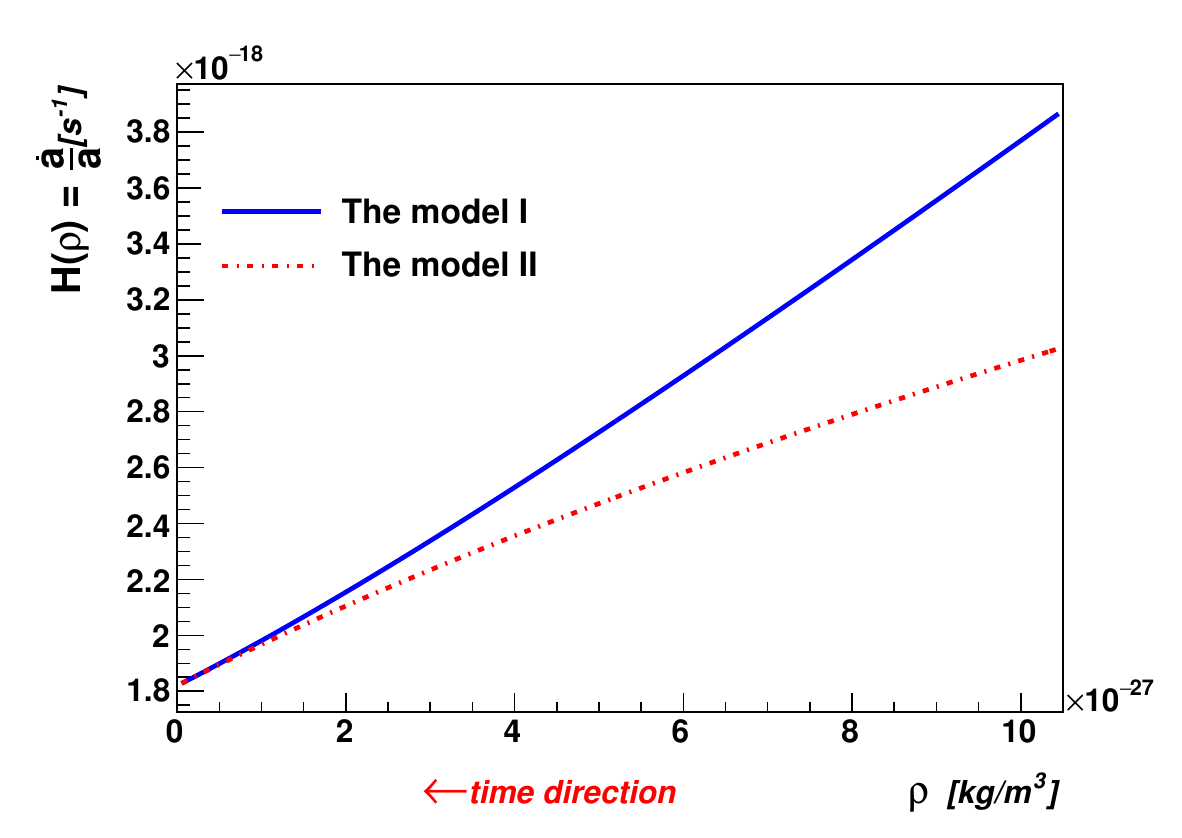}
\caption{\label{fig:5}\textit{This plot compares the behavour of $H(\rho)$ in two models I and II. Both functions decrease over time.}}
\end{center}
\label{H}		
	\end{figure}

\begin{figure}[h]
		\begin{center}
			\includegraphics[scale=0.43]{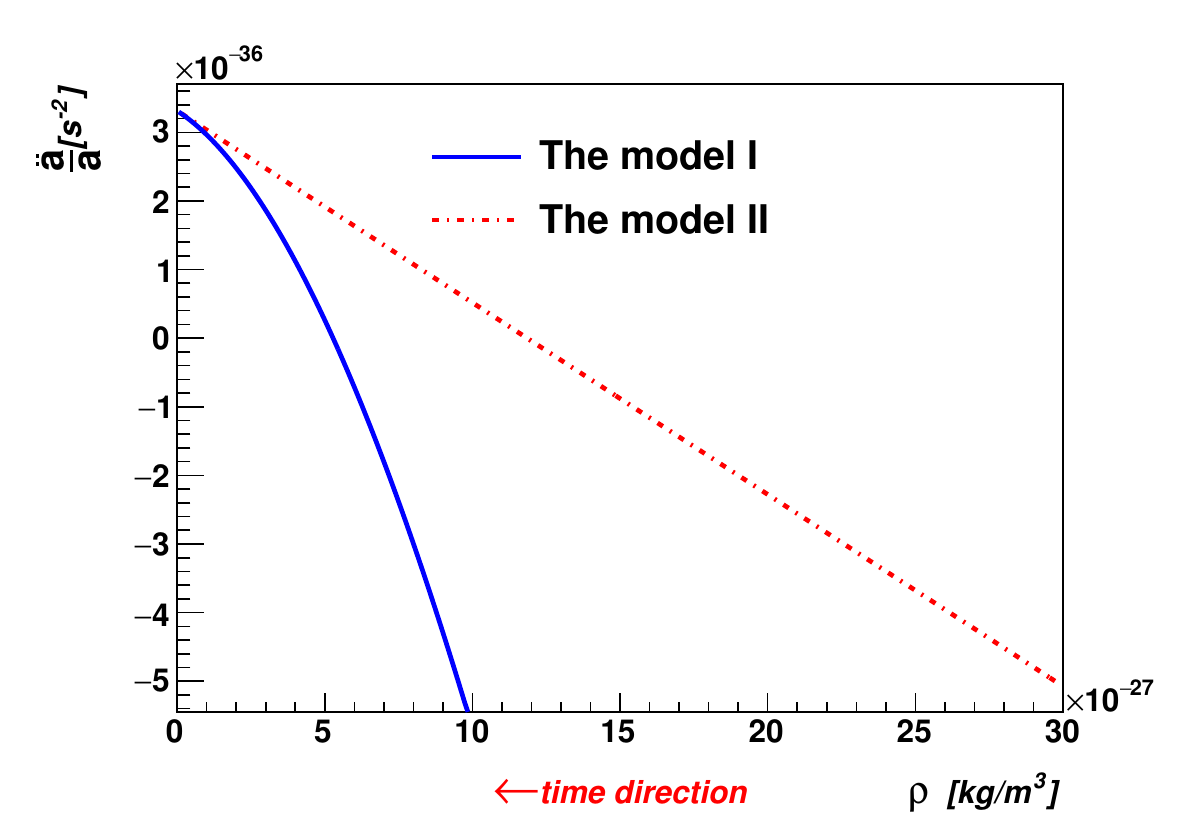}
\caption{\label{fig:6}\textit{This plot compares $\ddot{a}/a$ as function of $\rho $ in model I and model II. They both increase over time. That means the Universe in each of these models expands at an increasing acceleration.}}
\end{center}
\label{a}		
\end{figure}

Calculations similar to those for the  previous model immediately give  
\begin{align}
H(t)=&\left\lbrace \frac{c^2\Lambda \eta(1+\varepsilon)\left(4\Lambda+c^2k\rho \right)^\varepsilon }{3}\right.\nonumber\\
&\left. +\frac{c^4k\rho}{2}-\frac{c^4k\rho\eta(1+\varepsilon)\left(4\Lambda+c^2k\rho \right) ^\varepsilon}{6} \right.\nonumber\\
&\left. +\eta\varepsilon(1+\varepsilon)(4\Lambda+c^2k\rho)^{\varepsilon-1}c^4k\rho(\Lambda+c^2k\rho)\right.\nonumber\\
&\left. + \frac{c^2(4\Lambda+c^2k\rho)}{6}-\frac{c^2\eta\left(4\Lambda+c^2k\rho\right)^{1+\varepsilon}}{6} \right\rbrace^{1/2} 
\label{mk3}
\end{align}
and 
\begin{align}
3\left[1+\eta(1+\varepsilon)(4\Lambda +c^2k\rho)^\varepsilon \right] \frac{\ddot{a}}{a}=c^2\Lambda - 2c^4k\rho \nonumber\\
-3\left[ 1-2\eta(1+\varepsilon)(4\Lambda +c^2k\rho)^\varepsilon\right] H^2(t) \nonumber\\
-\frac{c^2\eta(4\Lambda+c^2k\rho)^{1+\varepsilon}}{2}+\frac{4c^2\Lambda + c^4k\rho}{2} \nonumber\\
-\frac{9\eta\varepsilon(1+\varepsilon)c^4k\rho(2\Lambda+3c^2k\rho)(4\Lambda+c^2k\rho)^{\varepsilon-1}}{2} \nonumber\\
-9\eta\varepsilon(\varepsilon^2-1)c^6k^2\rho^2(\Lambda+c^2k\rho)(4\Lambda+c^2k\rho)^{\varepsilon-2}. \label{mk4}
\end{align}
Choosing $v_{tg}\approx 200 - 300 km/s$, we get  \cite{Boehmer:2007kx}
\begin{equation} 
\varepsilon \approx 10^{-6}. \label{ep}
\end{equation} 
Thus, $H$ and $\dfrac{\ddot{a}}{a}$ as functions of $\rho$ are depicted in Fig. \ref{fig:3} and Fig. \ref{fig:4}, respectively.
These functions for models I and II are compared in Fig. \ref{fig:5} and Fig. \ref{fig:6}, respectively. A common feature of model I and II is that the acceleration of the Universe's expansion changes its sign at a given time ($t_2$), that is, the Universe goes from a deceleration phase to an acceleration phase. It's like a ball rolling up a hill (with a descreasing slope from bottom to top) to the top and then rolling down with an increasing slope from top to bottom.    
	
\subsection{Model III: $ f(R)=R-2\Lambda + \alpha R^2  +\displaystyle\frac{\gamma}{R} $}
\noindent
 
\begin{figure}[h]
		\begin{center}
			\includegraphics[scale=0.43]{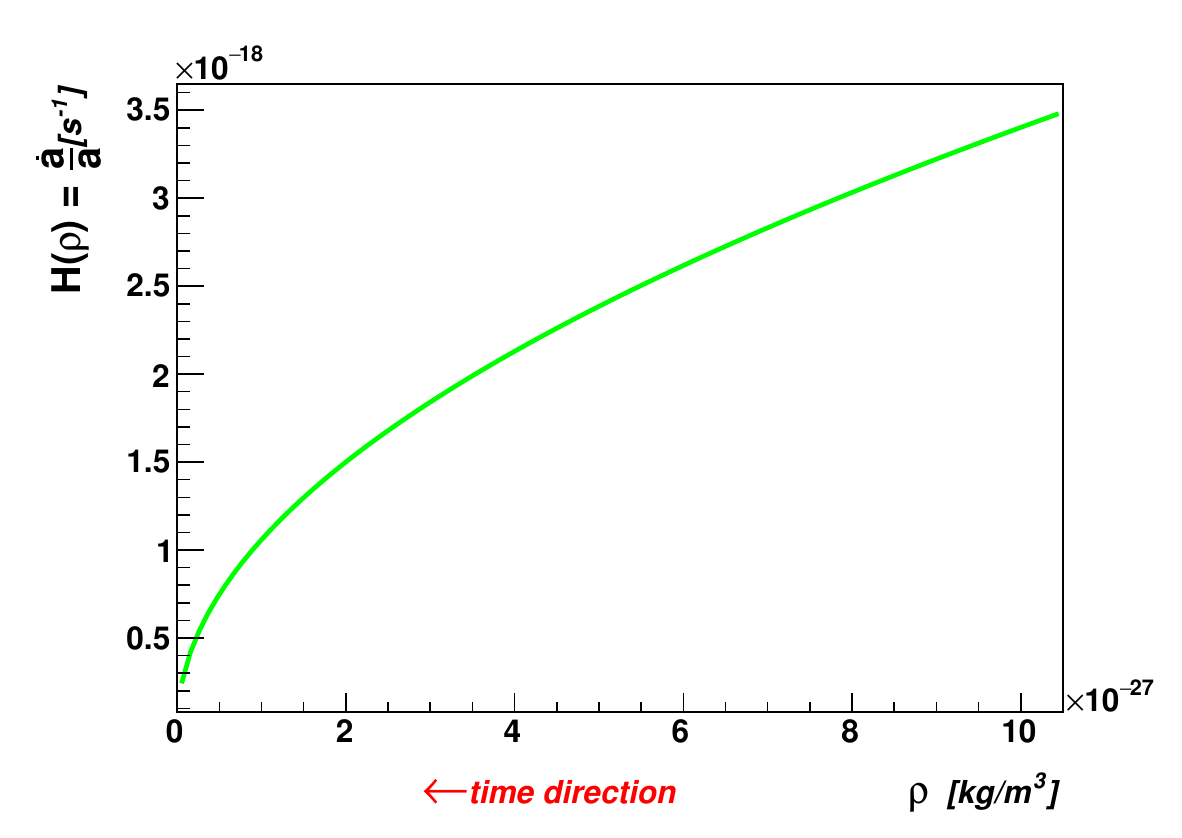}
\caption{\label{fig:7}: \textit{Plot of $H$ as a function of $ \rho $ in the model III. In this model, $H$ decays over time in contrast to those in the previous models I and II.}}
\end{center}
\label{Hiii}		
	\end{figure}

\begin{figure}[h]
		\begin{center}
			\includegraphics[scale=0.43]{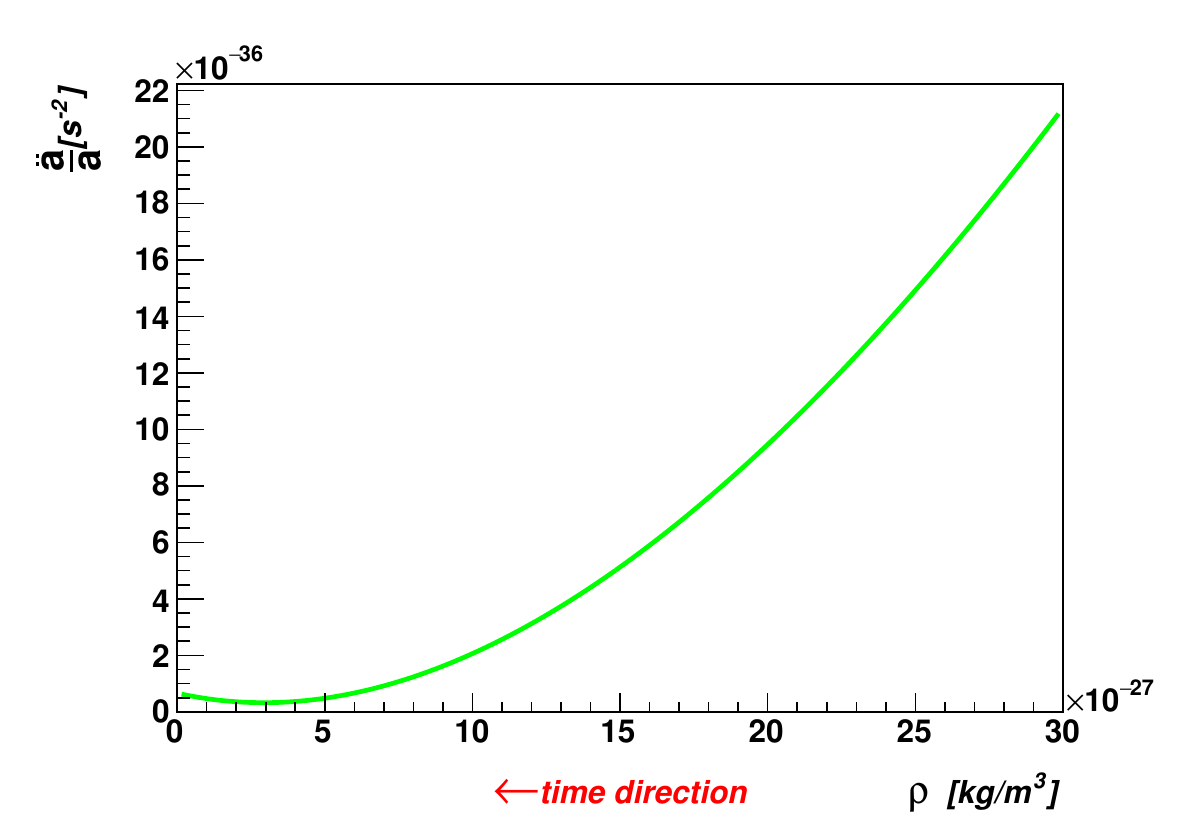}
\caption{\label{fig:8}: \textit{Plot of $\ddot{a}/a $ as a function of  $\rho $ in the model III. It  shows the time decay of $\ddot{a}/a$ in contrast to those increasing with time in the previous models I and II. It's similar to the acceleration of a ball rolling down a hill with a decreasing slope.
}}
\end{center}
\label{Hiii}		
	\end{figure}

 	The model-III with two terms $\alpha R^2$ (dominating in an early Universe) and $\gamma/R$ (dominating in a late Universe) can describe both the early and late Universe if $\alpha$ and  $\gamma$ are chosen appropriately.
 	This model resembles model-I (and, to some extent, similar to  the Starobinsky model) in first three terms  and, thus, can be used to describe an early Universe when the term $\gamma/R$ can be neglected. 
	For a late era of the Universe, 
	$ \alpha R^2 $ is small compared to $R$ and 
	$\displaystyle\frac{\gamma}{R} $
	we have 
	\begin{equation}
	f(R)\approx R-2\Lambda +\frac{\gamma}{R} \label{ga1}
	\end{equation}
	with   $\lambda h(R)= \displaystyle\frac{\gamma}{R} $. 
	Using  \eqref{H4}, \eqref{ro1a}, \eqref{ro1aa} and the approximations 
	\begin{align}
	&\overline{\Lambda}\simeq4\Lambda + c^2k\rho(t), \label{ga2}\\
	&\overline{\Lambda}^2 \simeq16\Lambda^2+8\Lambda c^2k\rho(t), \label{ga3}
	\end{align}
	we get the solution \eqref{H3} in the form (neglecting those terms of the second order $\lambda^2$ and higher orders of $\lambda$)
		\begin{align}
		 &H(t)=
		\nonumber\\& 	
        \sqrt{\frac{1}{3}c^2\Lambda +\frac{1}{3}c^4k\rho(t)-\frac{c^2\gamma}{16\Lambda}+\frac{c^4\gamma k\rho(t)}{16\Lambda^2}+\frac{c^6\gamma k^2\rho^2(t)}{384\Lambda^3}}, \label{ga5}
		\end{align}
	%
\begin{align}
&\left[6-\frac{3\gamma\left( 16\Lambda^2-8\Lambda c^2k\rho\right) }{256\Lambda^4} \right] \frac{\ddot{a}}{a}  
\nonumber\\&
-\left[3-\frac{6\gamma\left( 16\Lambda^2-8\Lambda c^2k\rho\right) }{256\Lambda^4} \right]H^2  
\nonumber\\&
+\frac{\gamma (\Lambda+ 2c^2k\rho)c^2}{8\Lambda^2}-\Lambda c^2=0. \label{gabs}
\end{align}
	Hence 
	%
\begin{align}
		&a(t) =a_0\exp{\int^t_{t_0}H(t)dt}.
        \label{ga6}
\end{align}
	%
	As  $ \rho(t)\longrightarrow 0 $ when $ t\longrightarrow \infty $ 
	(see \eqref{ga5})
	\begin{equation}
	H(\infty)=\sqrt{\frac{1}{3}c^2\Lambda -\frac{c^2\gamma}{16\Lambda}}, \label{ga7}
	\end{equation} 
	we see that a perturbation term  
	$ \displaystyle\frac{c^2\gamma}{16\Lambda} $ 
	still contributes to the GR.
	If $\gamma$ in \eqref{ga7} takes the value  $\gamma=\frac{16\Lambda^2}{3}$, then $H(\infty)=0$, and the Universe ceases to expand at $t\longrightarrow\infty$. This situation is consistent with the current commonly accepted  flatness of the Universe. This means that at some cosmic moment 
	the expansion of the Universe begins to slow down until the full stop at the infinitely distant future, as also shown by this model III (see Figs \ref{fig:7} and \ref{fig:8}). Models of this type were previously discussed in \cite{Nojiri:2003ft}, but the arbitrariness still remains there.\\		

	However, it is useful to make a side note that if the recent data  of the Planck's mission  \cite{DiValentino:2019qzk},  
telling us that the Universe might be closed, can be confirmed (this statement still requires confirmation by further observations) this model or/and the metric might have to be modified to adapt to the observation.\\


%
\section{Numerical discussions} 
%
\noindent 

Let us now consider perturbation conditions numerically. In the system of units SI  with $ c = 299792458 m/s $, $ G = 6.67259 \times 10^{-11} kg^{-1} m^3 s^{-2} $ \text{and} $ k = 2.0761154 \times 10^{-43}kg^{-1}m^{-1}s^2 $, the "would-be" observed value of the Hubble parameter $H(\tau)$ at a very distant future $t=\tau \rightarrow \infty$ (at which $\rho \rightarrow 0$) is 
	\begin{equation}
	H(\tau)=\sqrt{ \displaystyle \frac{c^2\Lambda}{3}} \sim 0.18199 \times 10^{-17} s^{-1}, \label{k39a}
	\end{equation}
	following the cosmological constant determined via the data given in 2018 by the Planck collaboration \cite {Planck:2018vyg} (see also wikipedia.org):
	\begin{equation}
	\Lambda \sim 1.1056 \times 10^{-52} m^{-2}. \label{k39b}
	\end{equation} 
	From the perturbation conditions  (see \eqref{vtf5})
	\begin{equation}
	\lambda h(R) \ll 2\Lambda, \label{dk1}
	\end{equation}
	we have
	\begin{equation}
	\lambda h(R) \ll 2.21 \times 10^{-52} m^{-2}. \label{k39c}
	\end{equation}
	As an illustration, let's consider the model $ f(R)=R-2\Lambda + \lambda R^2 $, then the perturbation condition \eqref{dk1} will be $ \lambda R^2 \ll 2\Lambda $, taking $ R\sim 4\Lambda $ (see \eqref{k8}), we get the perturbation condition for this model 
	\begin{equation}
	\lambda \ll \frac{1}{8\Lambda}. \label{k41}
	\end{equation}
	Using \eqref{k39b}, we have
	\begin{equation}
	\lambda \ll 1.1306 \times 10^{51} m^2. \label{k42}
	\end{equation}
	Now we  do some estimate calculations on \eqref{ro3}. Let $\lambda$ take the limit value \eqref{k42}. Based on this value we estimate the Hubble parameter at the time, say, $t_1$ (counted from the Big Bang), when the perturbation condition starts to be applicable, that is, 
	\begin{equation}
	\frac{1}{3}c^4k\rho(t_1)=\frac{3\lambda}{2}c^6k^2\rho^2(t_1), \label{tt1}
	\end{equation}
	or 
	\begin{equation}
	\rho(t_1)=\frac{2}{9\lambda c^2k}=1.05\times 10^{-26}kg/m^3. \label{tt2}
	\end{equation}  
	Then 
	\begin{equation}
	H(t_1)\sim 3.8796\times 10^{-18} s^{-1}, \label{tt3}
	\end{equation}
	therefore, 
	\begin{equation}
	t_1\simeq \frac{1}{H(t_1)}\sim 8.17 ~\mbox{billion years}. \label{tt4}
	\end{equation}
	The contribution of the perturbation term to the GR is quite significant at an earlier time before $t_1$ and negligible at a very late time. \\

	Next, we estimate the Hubble parameter at the time $t_2$ when the accelerated expansion of the Universe starts   \eqref{pttt2},    
		\begin{equation}
		c^2k\rho(t_2) = \frac{\sqrt{612\lambda^2\Lambda^2+156\lambda\Lambda+1}-1-2\lambda\Lambda}{38\lambda}, \label{tt3}
		\end{equation}
		or
		\begin{equation}
		\rho(t_2)=0.5280225 \times 10^{-26} kg/m^3.
		\end{equation}
		We calculate the GR value of the Hubble parameter at that time $t_2$, with $ c^2k\rho_{GR}(t_2)=2\Lambda $ (see \eqref{pttt2a})
		\begin{align}
		H_{GR}(t_2)&=\sqrt{\frac{1}{3}c^2\Lambda +\frac{1}{3}c^4k\rho_{GR}(t_2)}
		\nonumber\\&
		\sim 0.315216 \times 10^{-17} s^{-1}, \label{tt5}
		\end{align}
		corresponding to the Hubble time 
		\begin{equation}
		t_{2GR}\simeq \frac{1}{H_{GR}(t_2)}\sim 10.05 ~\mbox{billion years}. \label{tt5a}
		\end{equation}
		The Hubble parameter and time calculated according to  \eqref{ro3} are
		\begin{align}
		H(t_2)&=\sqrt{\frac{1}{3}c^2\Lambda +\frac{1}{3}c^4k\rho(t_2)+\frac{3\lambda}{2}c^6k^2\rho^2(t_2)}\nonumber\\
		&\sim 0.27827 \times 10^{-17} s^{-1}, \label{tt6}
		\end{align}
		and
		\begin{equation}
		t_2\simeq \frac{1}{H(t_2)}\sim 11.39 ~\mbox{billion years}, \label{tt6a}
		\end{equation}
		respectively (cf. Figs. \ref{fig:1} and \ref{fig:2}). The accelerated expansion of the Universe begins about 10.05 billion years after the Big Bang (according to the GR) or about 11.39 billion years after the Big Bang (according to the model-I, $f(R)=R-2\Lambda + \lambda R^2 $). The effect of the $f(R)$ theory is noticeable.
		The perturbation condition became applicable to the Model-I about 8.17 billion years after the Big Bang (i.e., about 5.6 billion years ago). 
		What about the model II with $f(R)=\eta R^{1+\varepsilon} -2\Lambda$, according to this model the acceleration of the Universe's expansion changes its sign at the time $t_2\sim 10.12 ~\mbox{billion years}$ (after the Big Bang) when $\rho(t_2)=1.1616\times 10^{-26}kg/m^3 $ and  $H(t_2)=0.313159\times 10^{-17} s^{-1}$ (see an  illustration in Fig. \ref{fig:4}).\\

	To finish we do a comparison of the perturbative solution of Model-I above with the solution of the Starobinsky model 
	$f(R)_{Starobinsky} = R + \displaystyle \frac {1}{6M^2} R^2$, 
	with a constant 
	$M$  \cite{DeFelice:2010aj,Starobinsky:1980te}.  Let us note again that the Model-I has a similar but not the same form with the Starobinsky model as the Lagrangian  $f(R)_{Starobinsky}$ does not have ${\cal L}_\Lambda=R-2\Lambda$ as a perturbative limit, unlike the Lagrangian $ f(R)=R-2\Lambda + \lambda R^2 $ of the Model-I  perturbatively developed around ${\cal L}_\Lambda$. We see first that 
	when considering the Universe in the late epoch (including the present time), as the term $R^2$ is too small (because $R$ is small), 
	the Starobinsky model gives almost no contribution to the Universe evolution in the present epoch, while the perturbative solution in the model-I gives a significant contribution for an appropriate value of $\lambda$. Therefore, the Starobinsky model, unlike the model-I, cannot describe well the late Universe but the early one. More precisely, the Starobinsky model contributes only during the era of the inflationary Universe, when $R$, that is, $R^2$ was very big. In the standard inflation model there must be  $\dot{H} <  0$, otherwise, the Universe could be  hyperinflationary \cite{DeFelice:2010aj}. 
	The Starobinsky model is consistent with the standard inflation condition for 
	$\dot{H}\propto \displaystyle\frac{-M^2}{6}$ \cite{DeFelice:2010aj}. 
	However, in the Starobinsky model the matter term (e.g., 
	$\rho(t)$) 
	has no contribution to the inflation rate. 
	The perturbative solution \eqref{ro5} of the model-I, when applied to the inflationary Universe, despite that the flat FLRW metric is used, as 
	$\dot{H} < 0$ 
	due to 
	$\dot{\rho}(t) < 0$, is consistent with the standard inflation, moreover, the matter term $\rho(t)$ also contributes to the inflation process of the Universe. 
	In other words, the model-I under the current perturbation approach, is more "flexible" than the Starobinsky model in accommodating different periods of the Universe's evolution. \\
	 
\section{Efective cosmological constant}

In fact, as will be seen, the effective cosmological constant is not a constant, but can be a parameter appearing as a time-varying function (even, if quantized, it can have a discrete range of values  \cite{CaoHNam:2022zbx}), which can be regulated accordingly. If in \eqref{H3} we use the notation  
\begin{align}
\Lambda_{eff}(\rho(t))= &{\Lambda}[1+\lambda h'(\overline{\Lambda})]-\frac{1}{2}k\rho \lambda h'(\overline{\Lambda})
   \nonumber\\& 
    -3\lambda \dot{h}'(\overline{\Lambda})\sqrt{\frac{\Lambda }{3}+\frac{k\rho}{3}}
	-\dfrac{1}{2}\lambda h(\overline{\Lambda}) \label{hsvt1}
\end{align}
and treat it as an effective cosmological constant (in \eqref{hsvt1} the terms of the second order $\lambda^2$ is neglected), 
where $\overline{\Lambda}=4\Lambda +k\rho$, 
then 
\begin{align}
H^2(t)=\frac{\Lambda_{eff}(\rho(t))+k\rho(t)}{3}. \label{hsvt2}
\end{align}
For the specific models considered above $\Lambda_{eff}(\rho(t))$ gets the following explicit forms:\\

- For the Model-I with $f(R)=R-2\Lambda +\lambda R^2$: 
\begin{align}
\Lambda_{eff}(\rho(t))=\Lambda +\frac{9\lambda}{2}k^2\rho^2(t). \label{hsvt3}
\end{align}

- For the Model-II with $f(R)=R -2\Lambda + \frac{\gamma}{R}$: 
\begin{align}
&\Lambda_{eff}(\rho(t))=
        \Lambda -\frac{3\gamma}{16\Lambda}+\frac{3\gamma k\rho(t)}{16\Lambda^2}+\frac{\gamma k^2\rho^2(t)}{128\Lambda^3}. \label{hsvt4}
\end{align}

%
- For the Model-III with $ f(R)=\eta R^{1+\varepsilon}-2\Lambda $:
\begin{align}
\Lambda_{eff}(\rho(t))=&\Lambda \eta(1+\varepsilon)\left(4\Lambda+k\rho \right)^\varepsilon \nonumber\\
&\left. +\frac{k\rho}{2}-\frac{k\rho\eta(1+\varepsilon)\left(4\Lambda+k\rho \right) ^\varepsilon}{2} \right.\nonumber\\
&\left. +3\eta\varepsilon(1+\varepsilon)(4\Lambda+k\rho)^{\varepsilon-1}k\rho(\Lambda+k\rho)\right.\nonumber\\
& + \frac{(4\Lambda+k\rho)}{2}-\frac{\eta(4\Lambda+k\rho)^{1+\varepsilon}}{2}.
\end{align}

Following \eqref{hsvt1} one can write the Lagrangian of a general perturbative $f(R)$-theory in the form
\begin{equation}
{\cal L}_G = R -2\Lambda_{eff}(\rho(t)). \label{hsvt5}
\end{equation}
Then it is easily to obtain all fundamental equations of this theory from the corresponding ones of the GR by replacing  the  cosmological constant  $\Lambda$ with the effective one $\Lambda_{eff}(\rho(t))$. Namely, starting from  the Lagrangian \eqref{hsvt5} we get the effective  Einstein equation 
\begin{align}
R_{\mu\nu}-\frac{1}{2}g_{\mu\nu}R +  g_{\mu\nu}\Lambda_{eff}(\rho(t))
=-kT_{\mu\nu}.\label{hsvt6}
\end{align}  
%
Furthermore, using \eqref{a39} and \eqref{hsvt6} we can easily obtain the effective Friedmann equations 
\begin{align}
&-3\frac{\dot{a}^2}{a^2} + \Lambda_{eff}(\rho(t)) =-k\rho, \label{hsvt7}\\
&-2\frac{\ddot{a}}{a}-\frac{\dot{a}^2}{a^2} +\Lambda_{eff}(\rho(t))=kP. \label{hsvt8}
\end{align}
Eq. \eqref{hsvt7} leads immediately to 
\eqref{hsvt2} which combined with \eqref{hsvt1} is nothing but \eqref{H3}. Similarly, we can see the equivalence betweet \eqref{hsvt8} and  \eqref{gt}.\\

All this shows the efficiency of the method used here and a perturbative $f(R)$-theory can be treated as an effective GR with the cosmological constant replaced by an effective one and thus, all original GR equations are replaced by corresponding effective ones. This simple procedure is possible thanks  to the perturbation approach. This procedure simplifies the process of  testing different models until finding a realistic one.	
%
	\section{Conclusions}
	\noindent 
	
	A general  perturbative solution of the $f(R)$ theory in the  FLRW cosmology is given in \eqref{a1} with $ H(t) $ given in \eqref{H3}. This  solution describes an  acceleratedly  expanding Universe ($\dot{a} > 0$ and $\ddot{a} > 0$). 
	In the literature there are a number of $f(R)$-modified models investigated but most of them are given with specific $f(R)$, rather than a general one, and so the corresponding solutions are specific and therefore they do not always easily fit the real evolution of the Universe. 
		The solution obtained here is general, though perturbative, with a general $f(R)$, is very convenient for testing any model, thus, it is not necessary to work separately on a specific model. In this way, one can try to test specific models one by one until getting a realistic  model. This procedure saves a lot of work in finding the right model (without the necessary to do, sometimes lengthy and tedious,  calculations for each test model). Furthermore, the cosmological constant is included in an effective cosmological parameter which can be regulated accordingly.  In the present paper, for illustration, we have demonstrated the application of this procedure to three specific models to see how to handle it. \\
	
	Applied to the model with  $f(R)=R-2\Lambda + \lambda R^2$ the general solution \eqref{H3} becoming  \eqref{ro3} and \eqref{ro4} shows  that this model can describe three consecutive eras of the Universe, 
	the solution of the model with $f(R)=\eta R^{1+\varepsilon}-2\Lambda$ is given in \eqref{mk3} and \eqref{mk4} which might solve the probelm of dark matter and dark energy, while for the model with  $f(R)=R-2\Lambda + \displaystyle \frac{\gamma}{R} $ the solution   is  \eqref{ga5} and \eqref{ga6}. We see that the Hubble parameter $ H $ depends on time  (while in the GR theory it is a constant). These  perturbative  solutions improve the GR solution in the sense that they show the evolution of H over time. We see that when $t\longrightarrow \infty$, the Hubble parameter of the model with  $f(R)=R-2\Lambda + \lambda R^2$ will approach that of  the GR theory (with $H=\sqrt{ \displaystyle \frac{c^2\Lambda}{3}}$), but  that in the models with $f(R)=\eta R^{1+\varepsilon}-2\Lambda$ and $f(R)=R-2\Lambda +  \displaystyle \frac{\gamma}{R}$ will not. The latter model, as a variant of model III in a late Universe, however, is consistent with an acceleratedly expanding and flat Universe (as is commonly accepted) for a given $\gamma$ (see comments by the end of Subsect. 3.3). At an early cosmic time, model III, approaching model-I, could describe the inflationary Universe (by adjusting its parameters). Model III, thus, could be a good candidate of a realistic cosmological model for both early and late cosmic epochs.\\

	We would like to stress that it is reasonable to work in the perturbative approach to the $f(R)$ theory, that means, we work with
	only those $f(R)$ satisfying the perturbation condition which could be present in some periods of the Universe evolution and have shown that a perturbative $f(R)$-theory could be a good theory of the Universe evolution in different stages. It is shown that the  perturbative $f(R)$-theory can be treated as an effective GR where the  cosmological constant is 
		$\Lambda$ replaced by an effective parameter $ \Lambda_{eff}[\rho(t)]$. This  treatment simplifies solving an $f(R)$-theory regardless its specific form.\\
	
	We could consider applying the present approach to an arbitrary (non flat in general) FLRW metric and the problem would become more complicated but we hope the method would work. This generalization is necessary if the statement \cite{DiValentino:2019qzk} on the recent data by Planck's satellite is confirmed.\\[1mm]
	{\bf Note added}: 
	After completing this 
	work we have been informed of the works  
	\cite{Nojiri:2003ft,Nojiri:2006gh,  Odintsov:2019evb,Nojiri:2019fft}  and related works, which exploit other aspects of the problem by other methods. 
\section*{Acknowledgement}
\noindent

	We would like to thank the referee for valuable  comments. This work is funded by the National Foundation for Science and Technology
	Development (NAFOSTED) of Vietnam under Grant
	No. 103.99-2020.50. \\	

\noindent
{\bf Data Availability Statement:} This manuscript has no associated data 722 or the data will not be deposited. [Authors' comment: This is a theoretical work and no experimental data].

\end{document}